\documentclass[11pt,letterpaper]{article}

\usepackage[T1]{fontenc}
\usepackage[utf8]{inputenc}
\usepackage{mathptmx}
\usepackage{textcomp}

\usepackage{geometry}
\usepackage{setspace}

\usepackage{amsmath}
\usepackage{graphicx}
\usepackage{color}
\usepackage{gensymb}
\usepackage{comment}
\usepackage{subcaption}
\usepackage{braket}
\usepackage{float}
\usepackage{authblk}
\usepackage{microtype}
\usepackage{array}
\usepackage[numbers,sort&compress]{natbib}

\title{ Effusivity-Controlled Interfacial Thermal Transport Revealed by Nanoscale Optical Thermometry}

\author{Adarsh B Vasista\thanks{Corresponding author: avasista@iiserb.ac.in}}

\author{Anita Kumari}

\author{Yash P. Mhaske}

\affil{Department of Physics, Indian Institute of Science Education and Research (IISER) Bhopal, India }

\date{}
\begin{document}

\maketitle

\begin{abstract}
Quantitative imaging of heat transport with high spatial and temporal resolution is essential for understanding thermal processes in heterogeneous systems, yet direct measurements of transient temperature fields at material interfaces remain challenging. Here, we employ time resolved thermal optical diffraction tomography (thermal ODT), a label-free nanoscale optical thermometry technique that reconstructs spatio-temporal evolution of three-dimensional temperature fields from thermally induced refractive-index changes. We show that thermal diffusion along an interface is controlled by their thermal effusivity contrast. We also derive an effective interfacial diffusivity that accurately describes the lateral propagation of thermal fields and validate the model through finite-element simulations across a broad range of liquid–glass interfaces. Surprisingly, liquids with lower bulk thermal diffusivities exhibit faster interfacial thermal spreading due to their lower effusivities. The measured diffusivities agree quantitatively with theoretical predictions over diverse material combinations. By combining volumetric thermal imaging with a general framework for interfacial heat transport, our work establishes thermal ODT as a powerful platform for investigating nanoscale thermodynamics and engineering heat flow in heterogeneous environments.
 
\end{abstract}

\section{Introduction}
Understanding and mapping thermal diffusion in micro- and nanoscale systems is crucial not only to address fundamental questions in nanoscale thermodynamics but also for a wide range of applications, including microfluidics\cite{9,10,11,12,13}, photochemistry\cite{14,15,16,17,18}, microbiology\cite{19,20}, and microelectronics\cite{21,22,23}. Most previous studies have focused on probing steady-state temperature distributions using either label-free techniques, such as quadriwave lateral shearing interferometry (QLSI)\cite{24} and optical diffraction tomography (ODT)\cite{25,2}, or label-based approaches including fluorescence intensity/anisotropy\cite{26,27,28} and Raman scattering\cite{29,31,32}. While there are a few reports on the measurement of heat diffusion from individual nanoscale objects in homogeneous media, a comprehensive understanding of thermal diffusion near heterogeneous planar interfaces remains largely unexplored\cite{33,34,35}. 

Thermal transport in most practical systems—including microelectronics, nanoelectromechanical systems (NEMS), etc —is governed by the presence of material interfaces. These interfaces reshape heat-flow pathways and give rise to transport behaviors that cannot be predicted from the properties of the constituent materials alone.  Of late, transient thermoreflectance microscopy has emerged as one of the most powerful methods for measuring thermal diffusion and probing interfacial heat transport with high temporal resolution\cite{36,37,38,39}. However, because it primarily probes temperature dynamics at or near reflective surfaces, investigating thermal transport across extended non/poorly reflective surfaces is challenging \cite{40}. Consequently, direct wide-field visualization of transient thermal diffusion across heterogeneous interfaces remains limited, hindering the development of general frameworks that connect bulk material properties to interfacial thermal transport.

An emerging alternative for studying thermal transport is thermal optical diffraction tomography (thermal ODT), which reconstructs temperature distributions from thermally induced refractive index variations\cite{2,41}. In the past, ODT has been extensively used for label-free, quantitative imaging of three-dimensional (3D) refractive-index distributions in biological systems\cite{41,45,46,47,48,49}, as well as for probing material anisotropy\cite{50} and other structural and optical properties of complex media. Unlike surface-sensitive approaches, thermal ODT provides volumetric and label-free imaging of the spatiotemporal evolution of temperature fields hence acting as a unique platform for quantifying thermal diffusivity and uncovering the physical parameters governing heat transport.

Thermal diffusivity is a fundamental material property that quantifies the rate at which heat propagates through a medium. Together with thermal conductivity, it plays a crucial role in the design and optimization of thermal insulators, conductors, thermoelectric devices, etc. In heterogeneous systems, heat transport is governed not only by thermal diffusivity but also by thermal effusivity, defined as $e=\kappa/\sqrt{D}$ where $\kappa$ is the thermal conductivity and $D$ is the thermal diffusivity respectively, which characterizes a material's ability to exchange heat with its surroundings\cite{42}. Materials with higher thermal effusivity can dissipate heat more effectively while materials with lower effusivities have larger thermal inertia. In addition, heat transport across an interface may be further influenced by the thermal boundary (Kapitza) resistance, which introduces a temperature discontinuity and impedes heat flow between dissimilar materials. As a result, the thermal evolution near an interface depends on the effusivity contrast and interfacial thermal resistance of the constituent materials. Despite the widespread occurrence of interfaces in natural and engineered systems, a general framework for predicting effective interfacial thermal transport, together with direct experimental validation, remains largely unexplored.

We combine numerical simulations and experiments to uncover the governing principles of thermal diffusion near liquid--glass interfaces. By studying systems with varying thermal effusivity contrasts, we establish a general scaling law that relates interfacial thermal diffusion to the thermal diffusivities and effusivities of the constituent materials. We validate this relationship experimentally using time-resolved pump--probe optical diffraction tomography, which enables label-free, wide-field imaging of transient temperature fields and direct observation of heat diffusion dynamics. The resulting framework quantitatively predicts effective interfacial thermal diffusion across diverse material combinations, filling a critical gap in the understanding of interfacial heat transport. More broadly, our work introduces a powerful volumetric thermometry platform for investigating heat flow in heterogeneous environments and provides design principles for controlling thermal transport in complex systems.

\section{Results and Discussion}
\subsection{Working principle}

    \begin{figure}[h!]
    \centering
    \includegraphics[width=\linewidth]{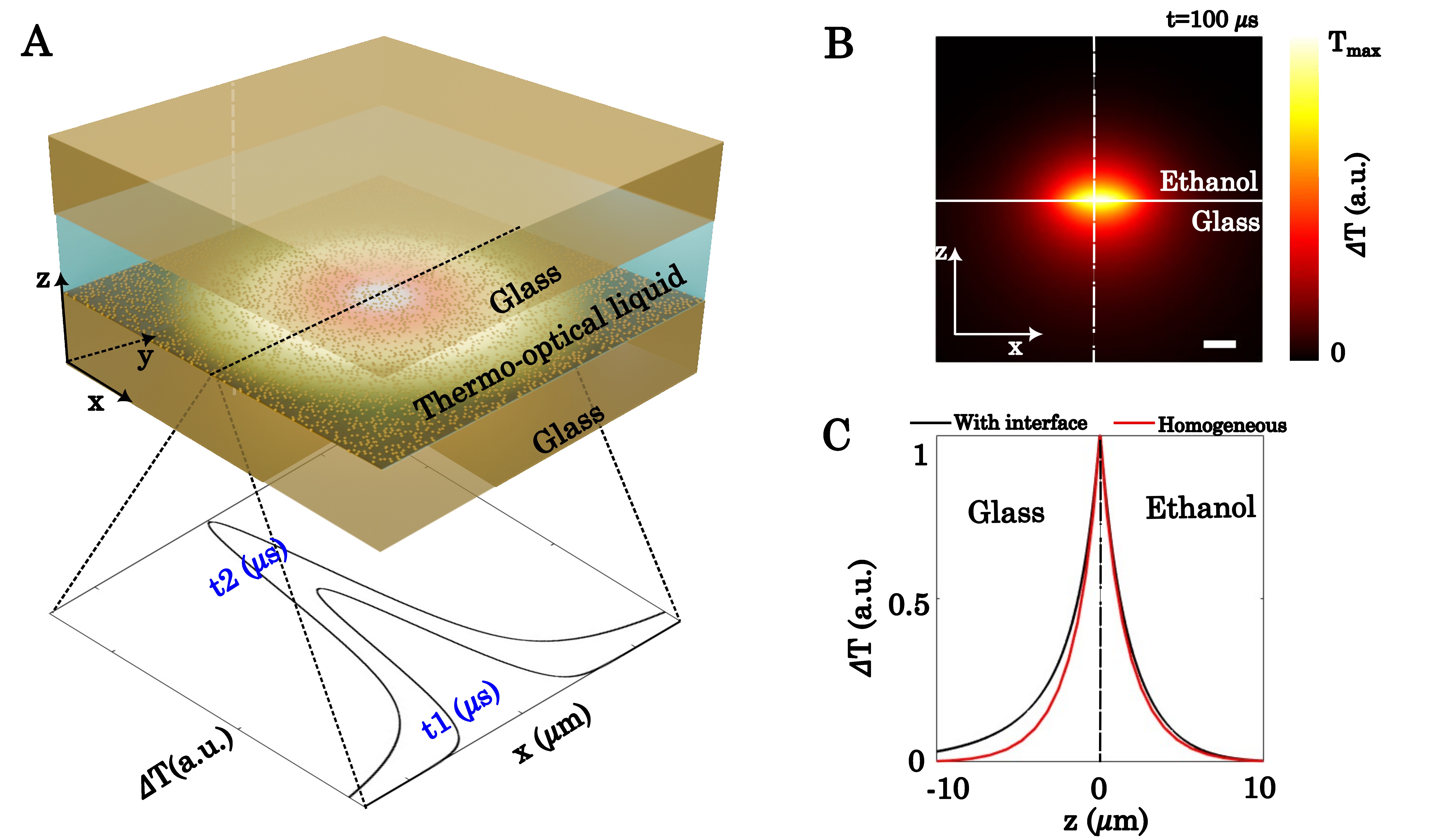}
    \caption{(A) Schematic of the experimental configuration used in this study. A micro-chamber was constructed using an Au NR functionalized glass substrate and an unocoated glass superstrate separated by a silicone spacer. The chamber was filled with different thermo-optical media (such as water, ethanol etc) with known refractive indices. Au NRs were excited by a focused 785 nm laser beam, which led to photo-thermal energy conversion and generated a transient thermal profile.(B) COMSOL simulated x-z snapshot of transient temperature distribution at t=100$\mu s$ for Gaussian heating (beam waist radius 1.5 $\mu m$) at the ethanol-glass interface (scale bar: 2$\mu m$). (C) Normalized temperature profile along x=0,(dash-dot line indicated in panel (B)) at t=100 $\mu s$. The profile for the glass-ethanol system is compared with that of a homogeneous ethanol medium.}
    \label{fig:Fig1}
\end{figure}

A schematic representation of the system under study is shown in Figure~\ref{fig:Fig1} (A). A microchamber was used as a model system to understand transient thermal behavior at nanoscale with Au nanorods (Au NRs) as nanoscale sources of heat by optically pumping at their absorption resonance. A typical microchamber consisted of a thermo-optical liquid (ethanol, water) sandwiched between a Au NR coated glass substrate and a glass superstrate. The thermal profile generated in the surrounding media by heating the Au NRs was measured in a time-resolved manner using pump-probe ODT \cite{2}. The detailed working principle of thermal ODT can be found elsewhere \cite{2,3}. In brief, thermal ODT relies on measuring thermally induced optical phase shifts of a probe beam under multiple illumination angles as it propagates through the microchamber. These angle-dependent phase profiles are combined according to the Fourier diffraction theorem to reconstruct a three-dimensional refractive-index distribution of the sample\cite{4}. The reconstructed refractive-index map is subsequently converted into a temperature distribution using the known thermo-optic coefficients of the surrounding media (see Methods for details). By providing label-free, volumetric snapshots of thermally induced refractive-index changes, thermal ODT enables direct visualization and quantitative measurement of transient thermal fields in micro- and nanoscale systems.

To demonstrate how interfaces modify thermal relaxation dynamics, we calculated the transient thermal response of a glass--ethanol--glass system using COMSOL. A Gaussian heat source located at the ethanol--glass interface was employed to mimic the spatial heating profile generated by the focused excitation beam in the experiments. Figure~\ref{fig:Fig1}(B) shows an axial ($x-z$) cross cut of temperature profile at t=100 $\mu s$ for Gaussian source with beam waist of radius 1.5 $\mu m$ at the ethanol-glass interface. In the case of planar media, the steady state thermal response is extensively studied and well understood \cite{5,6}, particularly the symmetric nature of the axial thermal profile even when the heat source is placed at the interface of two media with dissimilar thermal conductivities. However, the transient thermal profile doesn't follow the steady state response and shows asymmetry in the axial direction as shown in figure~\ref{fig:Fig1}(C). This asymmetry is significant at early time scales and becomes less significant as time evolves and the system moves towards the steady state (section S2, SI). Its important to note that the axial thermal profile across the interface mildly deviates from the homogeneous solution in the ethanol side, while a significant deviation appears on the glass side. 

\subsection{Evaluating thermal transients in heterogeneous systems}
If a thermal source with heat power $Q$ is placed, at the origin, in a homogeneous medium with thermal conductivity $\kappa$, diffusivity $D$, then temperature $T(r,t)$ at any spatial position $r$ and at any instant of time $t$ after the source is \textit{switched on} is given by \cite{7}
\begin{equation}
    T(r,t)=\frac{Q}{4\pi\kappa r} (1-erf(\frac{r}{\sqrt{4Dt}}))
    \label{eq1}
\end{equation}

where $erf()$ is the error function, which reduces to the well known $T(r)=\frac{Q}{4\pi\kappa r}$ profile in the steady state. If the heat source is placed at the interface of two planar media with thermal conductivities $\kappa_1$ and $\kappa_2$, then the steady state thermal profile in both media is given by \cite{5} 
\begin{equation*}
    T(r)=\frac{Q}{4\pi\kappa_{avg}r}
\end{equation*}
where $\kappa_{avg}$ is the average thermal conductivity of two media. 

\begin{figure}[h!]
    \centering
    \includegraphics[width=\linewidth]{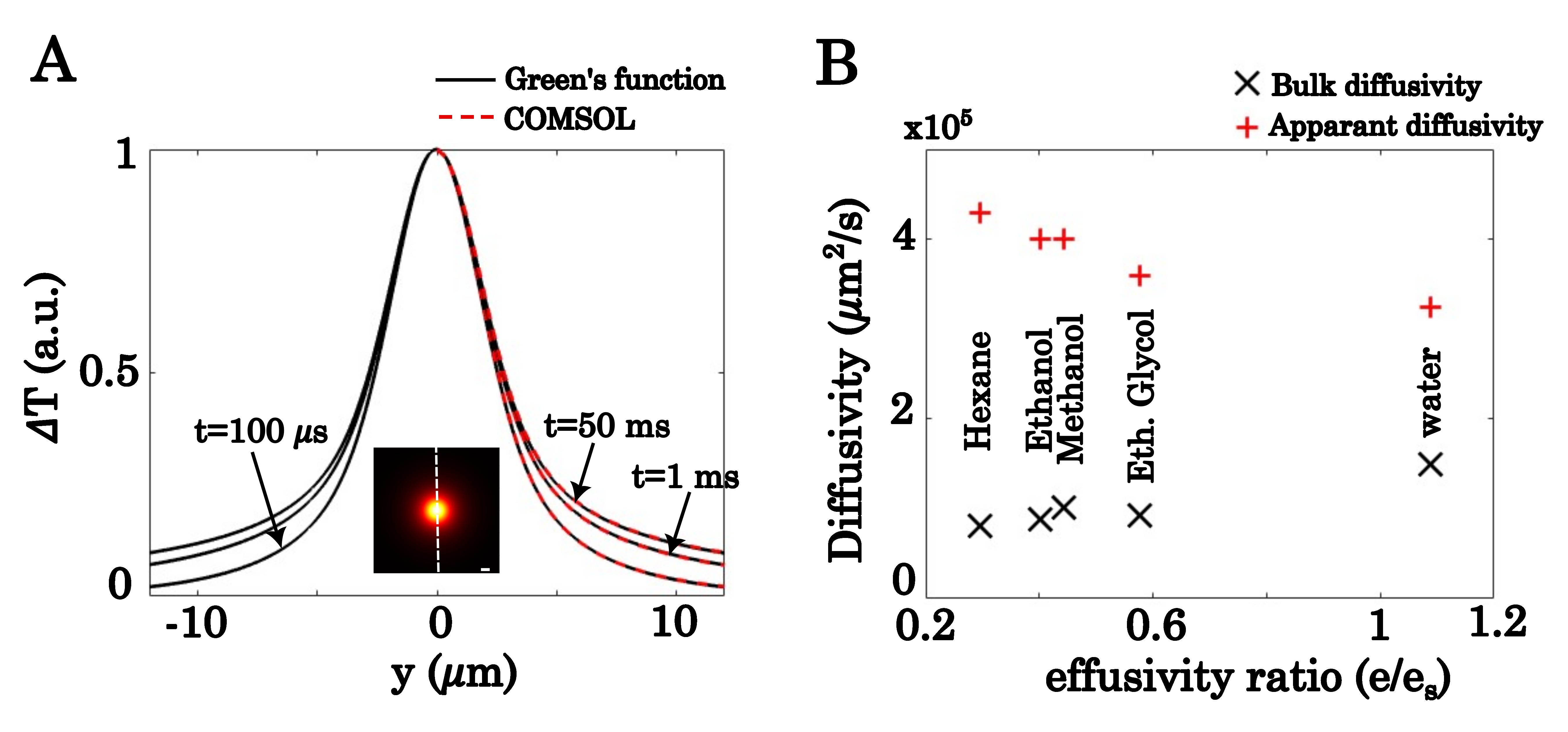}
    \caption{(A) Comparison of normalized temperature profiles along $x=0$ (dashed white line in the inset) at the ethanol--glass interface obtained from COMSOL simulations and a two-dimensional Green's function approximation using an apparent thermal diffusivity $D_{\mathrm{app}} = 4\times10^{5}\,\mu\mathrm{m}^{2}\mathrm{s}^{-1}$. Results are shown for $t=100\,\mu\mathrm{s}$, $1\,\mathrm{ms}$, and $50\,\mathrm{ms}$. (B) Apparent diffusivity ($D_{\mathrm{app}}$) and bulk diffusivity ($D_{\mathrm{bulk}}$) for different liquids as a function of the liquid-to-glass thermal effusivity ratio, illustrating the enhancement of interfacial heat spreading relative to bulk thermal diffusion.}
    \label{fig:Fig2}
\end{figure}

To understand thermal transport along an interface in a thermoplasmonic system, such as nanoheaters deposited on a glass substrate (Figure 1), the thermal effusivity ratio becomes a key parameter. In a homogeneous medium, the thermal diffusion length scales as $l \sim \sqrt{Dt}$. For diffusion along an interface, we introduce an effective diffusion length based on effusivity-weighted contributions from the two media, $l \sim \frac{e_1\sqrt{D_1 t}+e_2\sqrt{D_{2} t}}{e_1+e_2}$, where $e_1$ and $e_2$ are the thermal effusivities and $D_{1}$ and $D_{2}$ are the thermal diffusivities of the two materials forming the interface. This leads to an apparent interfacial thermal diffusion length through $l_{\mathrm{app}} \sim \sqrt{D_{\mathrm{app}} t}$ where

\begin{equation}
    D_{\mathrm{app}}= \left( \frac{e_1\sqrt{D_{1}}+e_2\sqrt{D_{2}}} {e_1+e_2}\right)^2
\end{equation}

Using this relation the temperature field can be modeled as diffusing within a homogeneous plane characterized by the apparent diffusivity, $D_{\mathrm{app}}$. This simplified description provides an intuitive framework for understanding interfacial thermal transport. Figure 2(A) shows the temperature profile along the ethanol–glass interface ($x=0$) calculated using COMSOL for Gaussian source localized at the interface. The resulting temperature distribution was overlaid with the two-dimensional Green's function solution (eq. 1), using an apparent diffusivity of $D_{\mathrm{app}}$ = 4 $\times$ 10$^5$$\mu\mathrm{m}^2/\mathrm{s}$ (calculated using eq. 2), showing an excellent agreement. Similar agreement was observed for a range of solvent–glass interfaces (Section S3, SI), validating the proposed effective diffusivity model.

Figure 2(B) compares the measured apparent diffusivities with the corresponding bulk diffusivities as a function of the effusivity ratio relative to glass. Interestingly, interfaces with lower effusivity ratios exhibit larger apparent diffusivities, even when the liquid itself possesses a lower bulk diffusivity. Physically, a low-effusivity medium is less effective at transporting heat away from the interface into its bulk. As a result, the thermal energy remains confined near the interface, where transport is increasingly governed by the higher-diffusivity glass substrate. This leads to a faster apparent spreading of the thermal field along the interface.

As the effusivities of the two materials become comparable, the interfacial diffusivity approaches the value expected from a conventional averaging of the thermal properties of the two media. This behavior may appear counterintuitive at first. For example, water possesses both higher thermal conductivity and higher thermal diffusivity than ethanol. Nevertheless, thermal diffusion along the ethanol–glass interface is faster than along the water–glass interface because the lower effusivity of ethanol confines heat closer to the interface, allowing the highly diffusive glass substrate to dominate the transport dynamics. Having established the theoretical and numerical description of thermal diffusion along interfaces, we next investigate this phenomenon experimentally.
\subsection{ODT maps the expansion of the thermal envelope}

  \begin{figure}[h!]
    \centering
    \includegraphics[width=\linewidth]{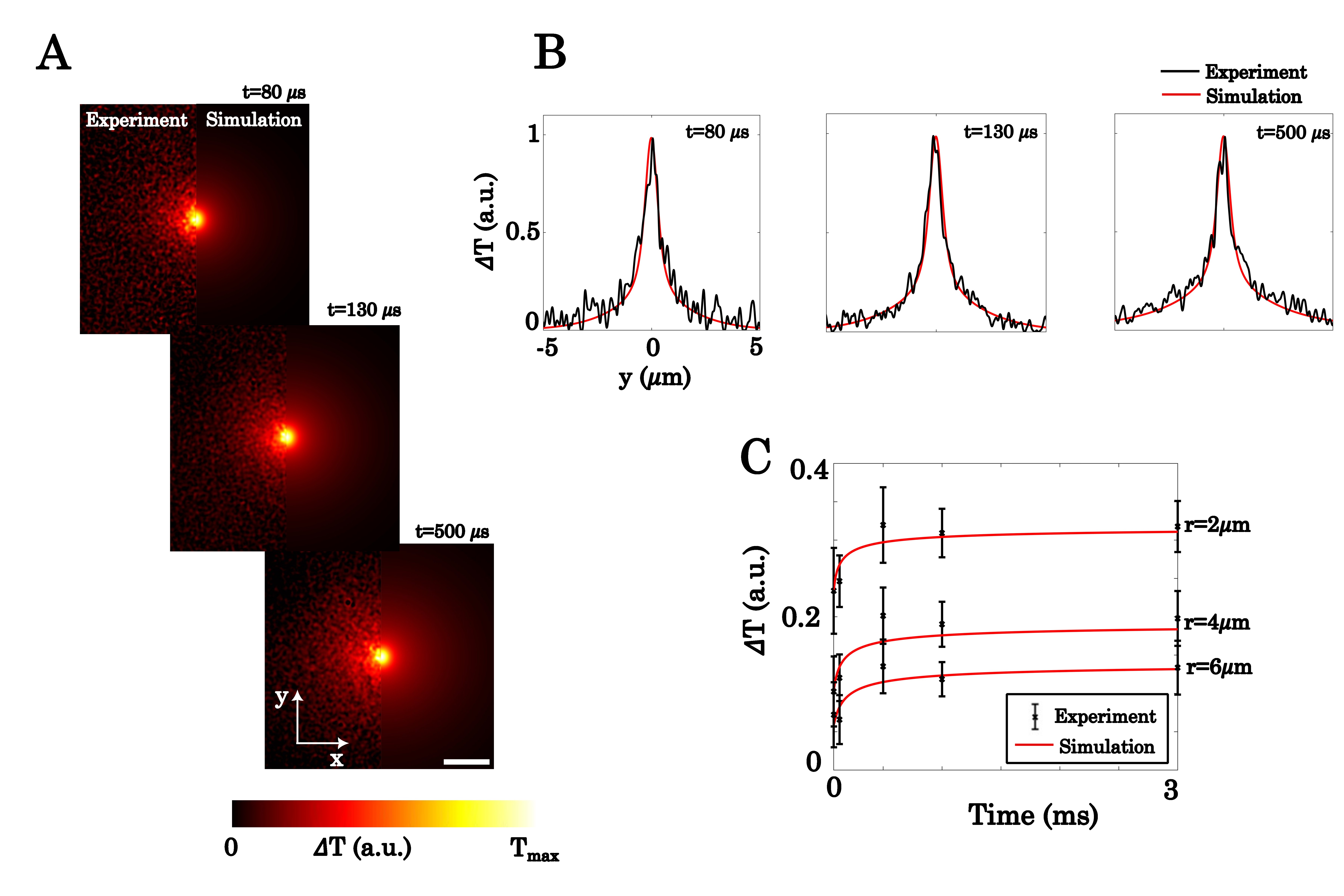}
    \caption{(A) Experimentally measured x-y thermal profiles when microchamber was filled with ethanol at t=80 $\mu$s, 130 $\mu$s, and 500 $\mu$s, compared with simulations obtained using the 2D Green's function, showing an excellent agreement (scale bar: 2 $\mu m$). (B) Normalized experimental thermal line profiles measured along x=0 for t=80 $\mu s$, 130 $\mu s$ and 500 $\mu s$. The temperature evolution was compared with results calculated using the 2D Green's function method to determine the thermal diffusivity parameter. (C) Temperature evolution along the circumference of a circle centered at the excitation spot with radii 2 $\mu$m, 4$\mu$m, and 6$\mu$m. The experimental data were fitted using the Green's function model to extract the diffusivity value.}
    \label{fig:Fig3}
\end{figure}

We experimentally measured the thermal transients in glass-ethanol-glass microchamber using time resolved pump-probe based thermal- ODT (see methods). Figure \ref{fig:Fig3}(A) shows experimentally measured thermal maps at various times (100 $\mu$s, 130 $\mu$s, and 500 $\mu$s) after the pump was \textit{switched on}. The measured profiles show an excellent match with the numerical simulations by 2D Green's function approach that considered apparent thermal diffusivity  of glass - ethanol interface ($D_{\mathrm{app}}$=4$\times$10$^{5}$$\mu\mathrm{m}^2/\mathrm{s}$). The experimentally measured normalized thermal profiles along $x=0$ were compared against the simulations for better visualization in figure \ref{fig:Fig3}(B). We extract the diffusivity data from the experimental measurements by finding a fit of the average of normalized temperature values at circles of different radii centered at the laser spot (|r|= 2 $\mu m$, 4 $\mu m$, and 6 $\mu m$) with numerical simulations performed using eq. \ref{eq1} as shown in figure \ref{fig:Fig3}(C). The extracted diffusivity value was found to be $D=3.8\times10^{5}\mu m^2/s$ which is in excellent agreement with the predicted value $D_{\mathrm{app}}$. It is important to note that the finite axial resolution introduces spatial averaging of the reconstructed temperature field; however, because the same reconstruction procedure is applied at all pump–probe delays and the thermal diffusion length is comparable to or larger than the axial resolution over the measured timescales, the temporal evolution of the interface temperature remains representative of the underlying heat transport dynamics. Also, the surface coverage of Au NRs were kept to a minimum so that it does not affect the thermal diffusion across the interface.\cite{24,25,2}

 \begin{figure}[h!]
    \centering
    \includegraphics[width=\linewidth]{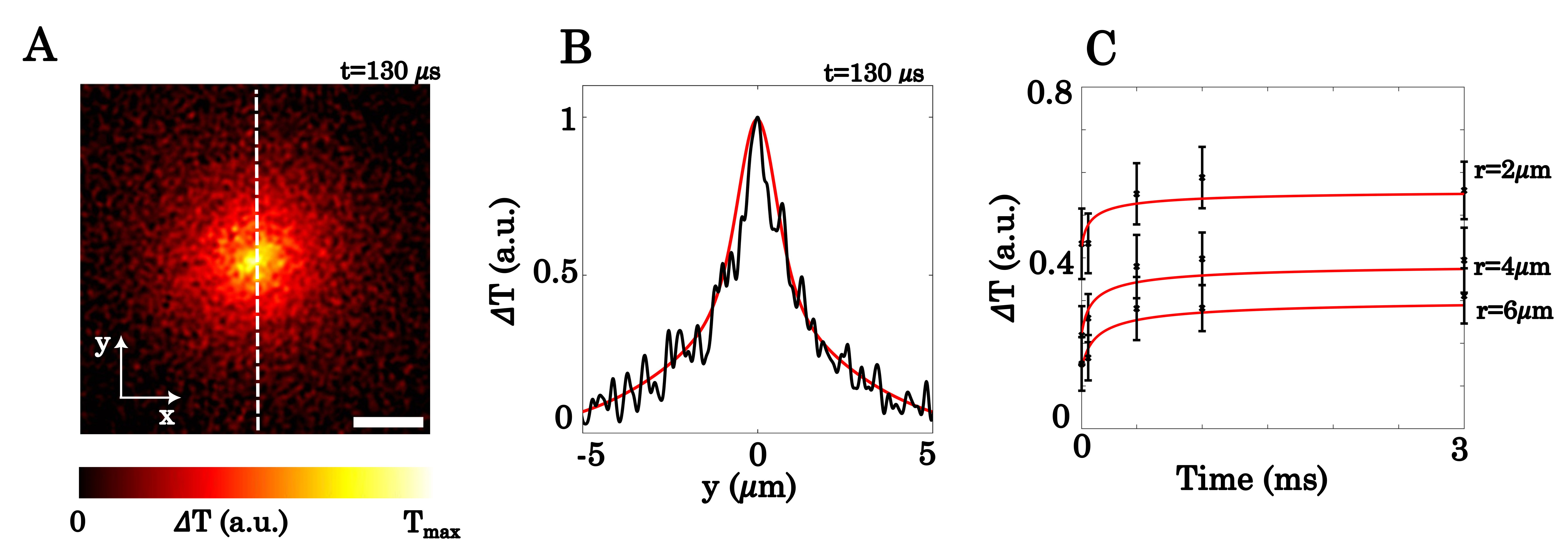}
    \caption{(A) Experimentally measured x-y thermal profile when the microchamber was filled with water at t=130 $\mu$s  (scale bar: 2 $\mu m$). (B) Normalized experimental thermal line profile along x=0 (across the dashed line in panel (A)) at t=130 $\mu$s compared with the thermal profile calculated with 2D Green's function approach. (C) Temperature evolution along the circumference of a circle centered at the excitation spot with radii 2 $\mu$m, 4$\mu$m, and 6$\mu$m. The experimental data were fitted using the Green's function model to extract the diffusivity value.}
    \label{fig:Fig4}
\end{figure}

To generalize our experimental observations, we performed time resolved temperature measurement for a glass-water-glass microchamber. Figure \ref{fig:Fig4} (A) shows the $x-y$ temperature map for t=130 $\mu$s and the normalized profile along $x=0$ is shown in figure \ref{fig:Fig4} (B). The line profile is superimposed with the numerically simulated temperature profile using eq.\ref{eq1} with $D_{\mathrm{app}}=3.25\times10^{5}\mu m^2/s$ (section S3, SI). The time evolution of the normalized temperature profile along the circumference of a circle with varying radii was used to extract the diffusivity value, in a similar manner as that of ethanol, as shown in figure \ref{fig:Fig4}(C). The extracted value of diffusivity was found to be $D_{\mathrm{app}}=3.08\times10^{5}\mu m^2/s$, again in close agreement with the predicted value. Also, the apparent diffusivities follow the inequality $D^{\mathrm{water}}_{\mathrm{app}}$ < $D^{\mathrm{eth}}_{\mathrm{app}}$ even though the bulk diffusivities of water is higher than that of ethanol, keeping the substrate parameters constant, clearly showing the effect of thermal effusivity mismatch in interfacial thermal transport.  

\section{Conclusion}

To summarize, we experimentally and numerically investigated the spatiotemporal evolution of temperature at the nanoscale using microchambers as model systems. Through numerical simulations and theoretical analysis, we demonstrate that for planar interfaces, the evolution of thermal profiles is governed not only by the thermal diffusivities of the constituent materials but also crucially by their thermal effusivities. In particular, we show that the expansion of the interfacial thermal envelope can be described by an effective interfacial diffusivity that emerges from the effusivity-weighted contributions of the adjoining materials. A striking consequence of this behavior is observed at glass–liquid interfaces. Even though the bulk thermal diffusivity and thermal conductivity of water are higher than those of ethanol, the thermal envelope at the water–glass interface evolves more slowly than at the ethanol–glass interface.

Using time-resolved pump–probe thermal optical diffraction tomography (thermal-ODT), we demonstrate that wide-area optothermal measurements can quantitatively extract interfacial thermal spreading dynamics and effective diffusivities in heterogeneous systems. The experimentally measured thermal evolution across multiple material interfaces shows excellent agreement with our numerical predictions, establishing thermal-ODT as a powerful label-free platform for probing transient heat transport at complex interfaces. Beyond reconstructing steady-state temperature distributions, our approach enables direct investigation of local thermal dynamics with temporal resolutions on the order of tens of microseconds. We anticipate that these capabilities will provide new opportunities for studying nanoscale thermodynamics and interfacial heat transport, while also offering valuable design and characterization tools for materials and devices used in thermal management, energy technologies, microfluidics, and optothermal engineering.

\section{Methods}
\subsection{Sample preparation}
Au NRs were synthesized using the method described elsewhere\cite{43}. A microchamber was prepared by placing silicon gaskets of 120 $\mu$m thickness on the Au NR coated glass substrate. The gap in the silicon gasket was filled with $\sim$ 20 $\mu$l of a thermo-optical liquid (water or ethanol), and the chamber was closed with an uncoated glass superstrate. 

\subsection{Time resolved thermal-ODT}
Time-resolved thermal optical diffraction tomography (thermal ODT) was employed to reconstruct the three-dimensional spatiotemporal temperature field. For each pump–probe delay, complex optical fields (amplitude and phase) were measured under multiple illumination angles using the off-axis holographic microscope described in Section S1, SI. The recorded holograms were Fourier transformed, and the first-order diffraction term was isolated in the spatial-frequency ($k$) space, frequency demodulated, and inverse Fourier transformed to retrieve the complex electric field. To improve phase sensitivity and enhance the signal-to-noise ratio, 100 pump-ON and pump-OFF measurements were acquired and averaged for each illumination angle. The scattered field, ($U_{\mathrm{scatt}}$) was then calculated within the Rytov approximation from complex fields measured with ($U_{\mathrm{obj}}$) and without ($U_{\mathrm{back}}$) optical heating using the relation $U_{\mathrm{scatt}}$=ln$(\frac{U_{\mathrm{obj}}}{U_{\mathrm{back}}})$. Repeating this procedure for all illumination angles enabled the construction of the three-dimensional Fourier space according to the Fourier diffraction theorem, from which the refractive-index distribution was reconstructed via inverse Fourier transformation. The reconstructed refractive-index maps were subsequently converted into temperature distributions using the calibrated temperature-dependent refractive-index relation of water/ethanol (see section S1, SI). To ensure reproducibility and minimize measurement inconsistencies, the entire experiment was repeated three times at the same location, and the resulting temperature maps were averaged. By performing these measurements at different pump–probe delays (0, 50, 420, 920, and 2920 $\mu$s), a series of three-dimensional thermal maps was obtained, enabling direct visualization of the transient evolution of heat diffusion. The exposure time of the camera was kept at 80 $\mu$s, which limited the temporal resolution of the microscope.

\subsection{COMSOL simulations}
2D axisymmetric finite element method numerical simulations were performed using COMSOL multiphysics. The substrate, superstrate, and thermo-optical material thickness was kept at 120 $\mu$m and the lateral simulation width was 800 $\mu$m. Natural convection boundary conditions were applied at the top and bottom surfaces of the microchamber, while all other boundaries were maintained at room temperature to mimic the experimental conditions. The different thermal parameters of materials (thermal conductivity, diffusivity etc ) are enumerated in section S2, SI. The pump laser was modeled as a Gaussian beam with a beam waist (BW) of 1.5$\mu m$ at the glass–liquid interface. 

\subsection{Data fitting using 2D Green's function method}
The apparent thermal diffusivity was determined by fitting the measured temperature distributions with a two-dimensional Green's function solution of the heat diffusion equation (eq. 1). To account for the finite camera exposure time (80 $\mu$s), the theoretical temperature profile was evaluated at a series of 20 time points within the exposure window for each pump--probe delay. These profiles were subsequently averaged to obtain the effective temperature distribution recorded by the camera. The apparent diffusivity (D$_{\mathrm{app}}$) was treated as the fitting parameter, and its value was obtained by minimizing the least-squares error between the experimentally measured and theoretically predicted temperature profiles. The fitting procedure was repeated for all pump--probe delay times analyzed in this work.







\section*{Acknowledgement}

ABV thanks IISER Bhopal for the research initiation grant and Anusandhan National Research Foundation (ANRF), Govt. of India for the Prime Minister's Early Career Research Grant (ANRF/ECRG/2024/006239/PMS). Authors thank Falko Schmidt for valuable suggestions.


\section{Author Contributions}
ABV built the experimental setup, data acquisition and analysis software, performed the experiment, and COMSOL simulations. YPM performed numerical simulations using the 2D Green's function method. AK prepared the Au NRs used in the experiments. ABV wrote the paper with inputs from AK and YPM. 

\section{Data Availability Statement}
All data generated or analyzed during this study are included in the
article and its Supplementary Information. 



\bibliographystyle{naturemag}
\bibliography{ref}

\clearpage

\setcounter{secnumdepth}{3}
\setcounter{section}{0}
\setcounter{figure}{0}
\setcounter{equation}{0}
\renewcommand{\thesection}{S\arabic{section}}
\renewcommand{\thesubsection}{S\arabic{section}.\arabic{subsection}}
\renewcommand{\thefigure}{S\arabic{figure}}
\renewcommand{\theequation}{S\arabic{equation}}

\begin{center}
  {\large\bfseries Supplementary Information\par}
  \vspace{0.5em}
  {\large Effusivity-Controlled Interfacial Thermal Transport Revealed by Nanoscale Optical Thermometry\par}
\end{center}
\vspace{1em}

\section{Experimental setup}

\begin{figure}[htbp]
  \centering
  \includegraphics[width=0.95\textwidth]{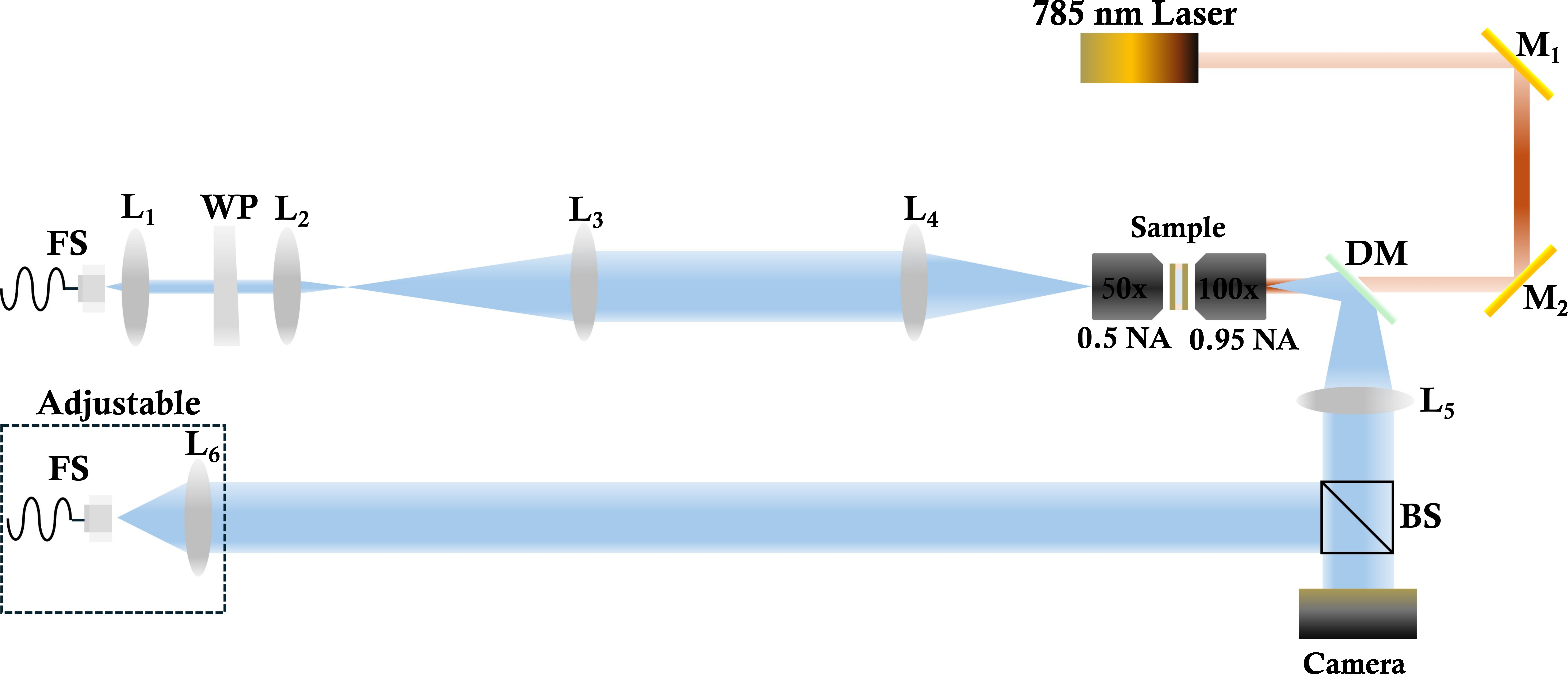}
  \caption{Schematic of the pump-probe time resolved thermal ODT setup. L: Lens, M: Mirror, BS: Beam splitter, DM: Dichroic mirror, WP: Wedge prism, FS: Fiber source}
  \label{fig:wgm_field}
\end{figure}

Figure S1 shows a schematic of the off-axis holographic microscope used to measure amplitude and phase maps in pump-probe configuration. The probe laser (457 nm) was split into reference and object beams using a fiber beam splitter. In the object path, the probe laser was focused onto the back aperture of the objective lens (50x, 0.5NA) to generate a wide field illumination using a combination of lenses $L_2$ ($f_{L_2}$=30 mm), $L_3$ ($f_{L_3}$=150 mm), and $L_4$ ($f_{L_4}$=150 mm). The probe light was collected in transmission configuration by a 100x 0.95 NA objective lens and the collected light was projected onto the camera using lens $L_5$ ($f_{L_5}$=200 mm) creating a magnification of 111 ( $f_{L_5}/f_{objective}$ ). The angle of illumination at the sample plane was controlled by the wedge prism $WP$. The reference beam was projected onto the camera at a small angle with respect to the optic axis of the microscope. The path length of the reference beam was adjusted by placing the fiber source module on an adjustable stage so as to match that of the object path. The Au NRs were excited using a pump beam of wavelength 785 nm through 100x objective lens. An NI-DAQ card was used to synchronize the pump and probe lasers to the camera acquisition. The frame rate of the camera was set at 40Hz for all experiments and the pump pulse duration was set at 5 ms. The data analysis pipeline including iterative regularization, phase sensitivity etc is detailed in ref\cite{2}. 

For glass--water--glass microchamber, the three-dimensional refractive index distribution was converted into a three-dimensional temperature map using the empirical relation

\begin{equation}
n(T)=b_0+b_1T+b_2T^2+b_3T^3+b_4T^4,
\end{equation}

where $T$ is the temperature and the coefficients are given by Ref.~\cite{44}:
$b_0=1.34359$,
$b_1=-1.0514\times10^{-4}$,
$b_2=-1.5692\times10^{-6}$,
$b_3=5.7538\times10^{-9}$, and
$b_4=-1.2873\times10^{-11}$.

For the glass--ethanol--glass microchamber, we used a linear relation between the refractive index and temperature given by
\begin{equation}
\Delta n(T)= \frac{dn}{dT}T
\end{equation}
 with $\frac{dn}{dT}$=-3.9$\times10^{-3}$ 
\section{COMSOL simulations}

The table below summarizes the thermal parameters of the materials used in the COMSOL simulations.\\
\begin{center}
\begin{tabular}{| m{3cm}| m{2cm}| m{2cm}| m{2cm}| } 
  \hline
  \textbf{Material} & \textbf{$\kappa(W/mK)$} & \textbf{$\rho(kg/m^3)$} & \textbf{$C_p(J/kgK)$} \\ 
  \hline
   Glass & 1.1 & 2210 & 830 \\ 
  \hline
  Water & 0.6 & 997 & 4000 \\ 
  \hline
Ethanol & 0.17 & 789 & 2440 \\ 
  \hline
Methanol & 0.2 & 786 & 2510 \\ 
  \hline
Hexane & 0.12 & 655 & 2260 \\ 
  \hline
Ethylene Glycol & 0.25 & 1110 & 2415 \\ 
  \hline
\end{tabular}
\end{center}
\subsection{Influence of the interface in thermal diffusion across boundaries}
\begin{figure}[ht]
  \centering
  \includegraphics[width=0.75\textwidth]{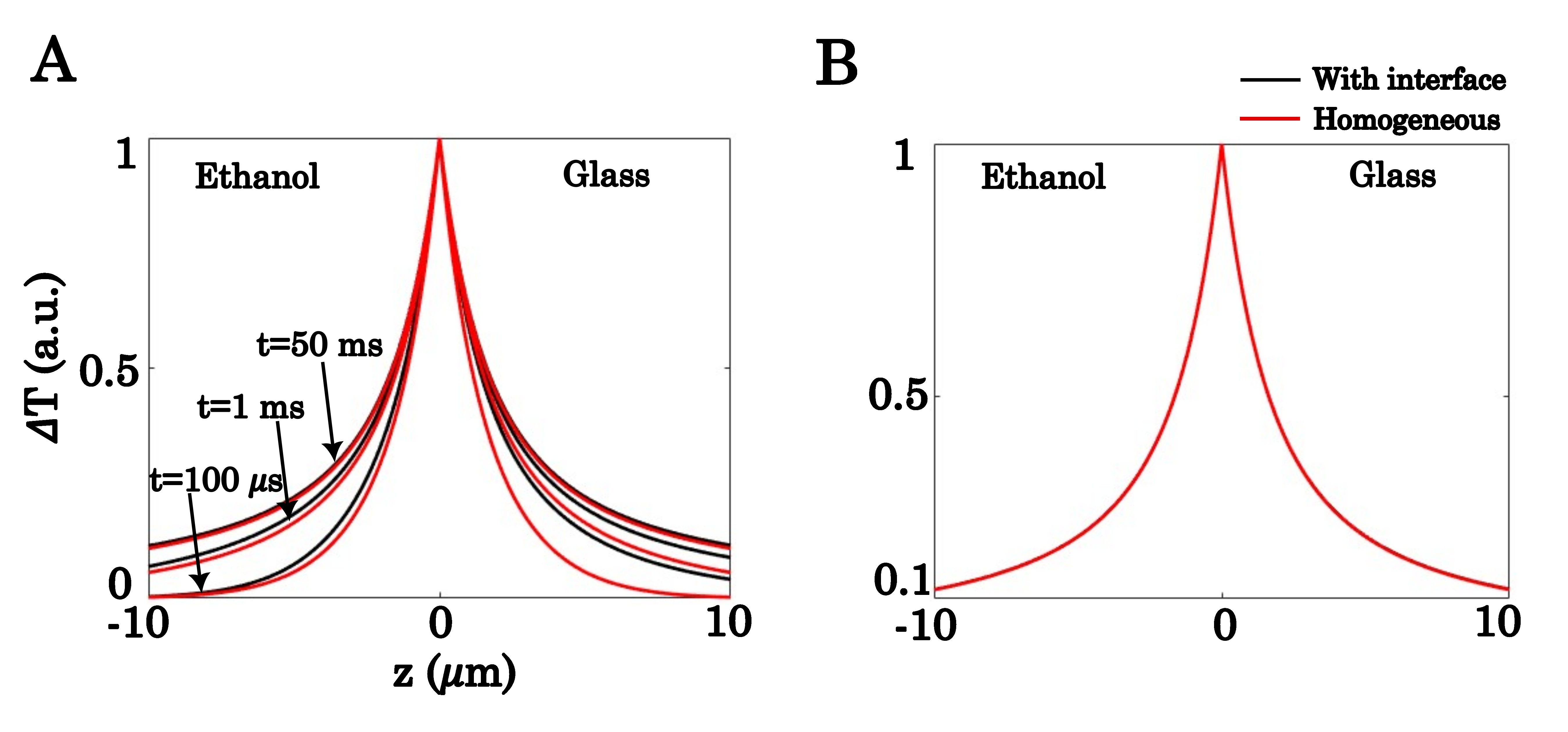}
  \caption{(A) Normalized axial temperature profiles along $(x,y)=0$ at $t = 100~\mu\mathrm{s}$, $1~\mathrm{ms}$, and $50~\mathrm{ms}$ for the glass--ethanol interface, compared with those in a homogeneous ethanol medium. At early times, the presence of the interface introduces asymmetry in the thermal profile; however, as time progresses, the profile becomes increasingly symmetric and gradually converges toward that of the homogeneous ethanol case. (B) Normalized axial temperature profiles along $(x,y)=0$ at steady state for the glass--ethanol interface and a homogeneous ethanol medium. At steady state, the thermal profile in the presence of the interface exactly matches that of the homogeneous medium}
  \label{S2}
\end{figure}

Figure \ref{S2}(A) shows the temporal evolution of the axial temperature profile for a glass--ethanol interface, together with the corresponding profile for a homogeneous ethanol medium. At early times, the interface introduces a pronounced asymmetry in the thermal distribution. On the ethanol side, the temperature profile mildly deviates from that of the homogeneous medium, whereas on the glass side significant deviations are observed due to the differing thermal properties of the two materials. As time progresses, the influence of the interface gradually diminishes, and the temperature profile becomes increasingly symmetric. At steady state, the profile approaches that of the homogeneous ethanol case, as shown in Fig.~\ref{S2}(B).

\section{Apparent diffusivities in multiple liquid-glass interfaces}
\begin{figure}[ht]
  \centering
  \includegraphics[width=0.75\textwidth]{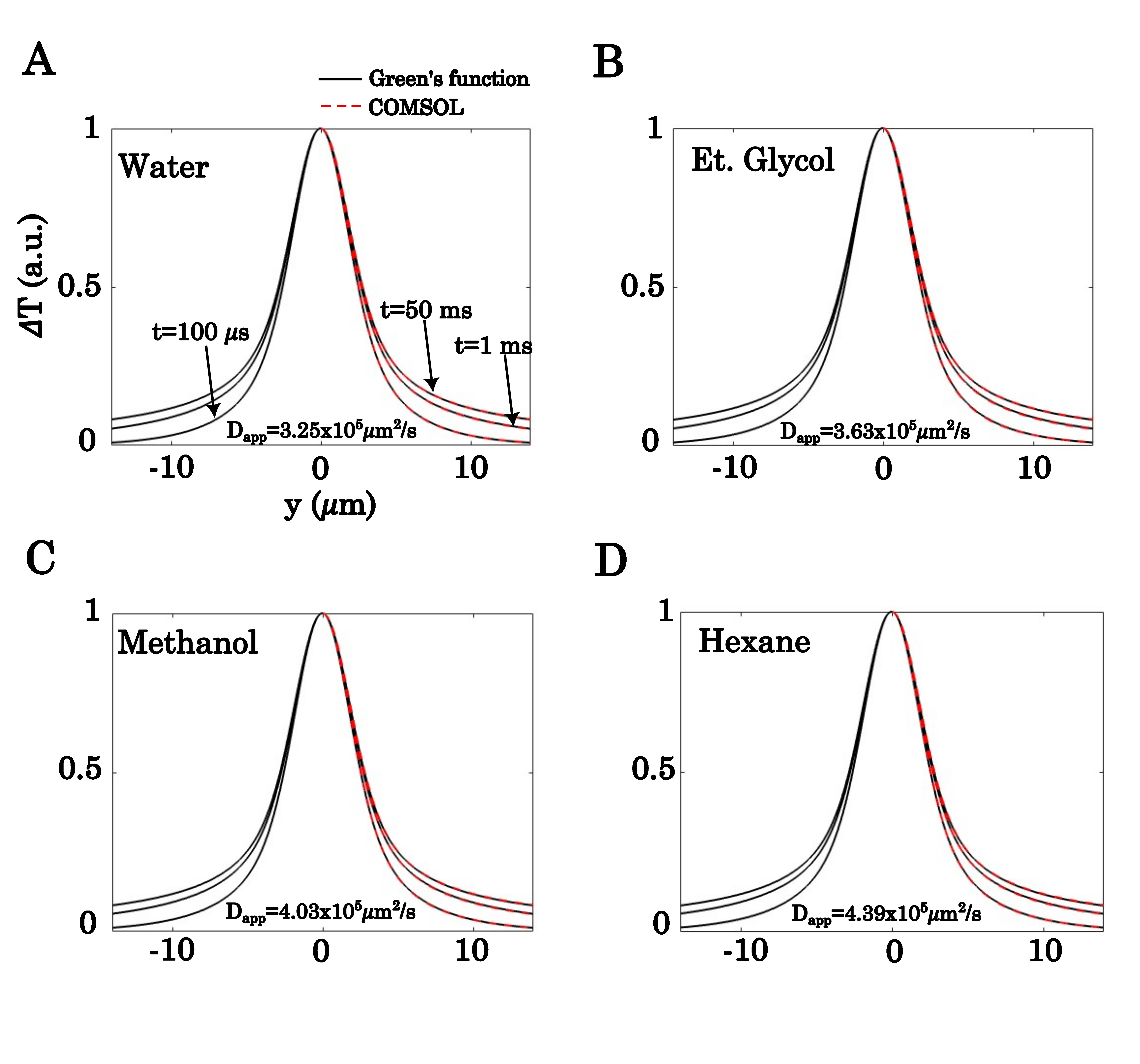}
  \caption{Comparison of normalized temperature profiles along $x=0$ (dashed white line in the inset) at the (A) Water--glass interface (B) Ethylene Glycol--glass interface (C) Methanol--glass interface and (D) Hexane--glass interface obtained from COMSOL simulations and a two-dimensional Green's function approximation respectively. Results are shown for $t=100\,\mu\mathrm{s}$, $1\,\mathrm{ms}$, and $50\,\mathrm{ms}$. }
  \label{S3}
\end{figure}

Figure~\ref{S3} presents the spatiotemporal evolution of thermal profiles at different liquid--glass interfaces obtained from COMSOL simulations. The corresponding profiles calculated using the 2D Green's function formalism, with the apparent diffusivity values ($D_{\mathrm{app}}$) shown in the inset, are overlaid for comparison. The apparent diffusivities were determined using Eq.~2 of the main manuscript. The excellent agreement between the Green's function model and the COMSOL results demonstrates that the proposed analytical framework accurately captures the thermal diffusion dynamics near liquid--glass interfaces.

\begin{figure}[ht]
  \centering
  \includegraphics[width=0.5\textwidth]{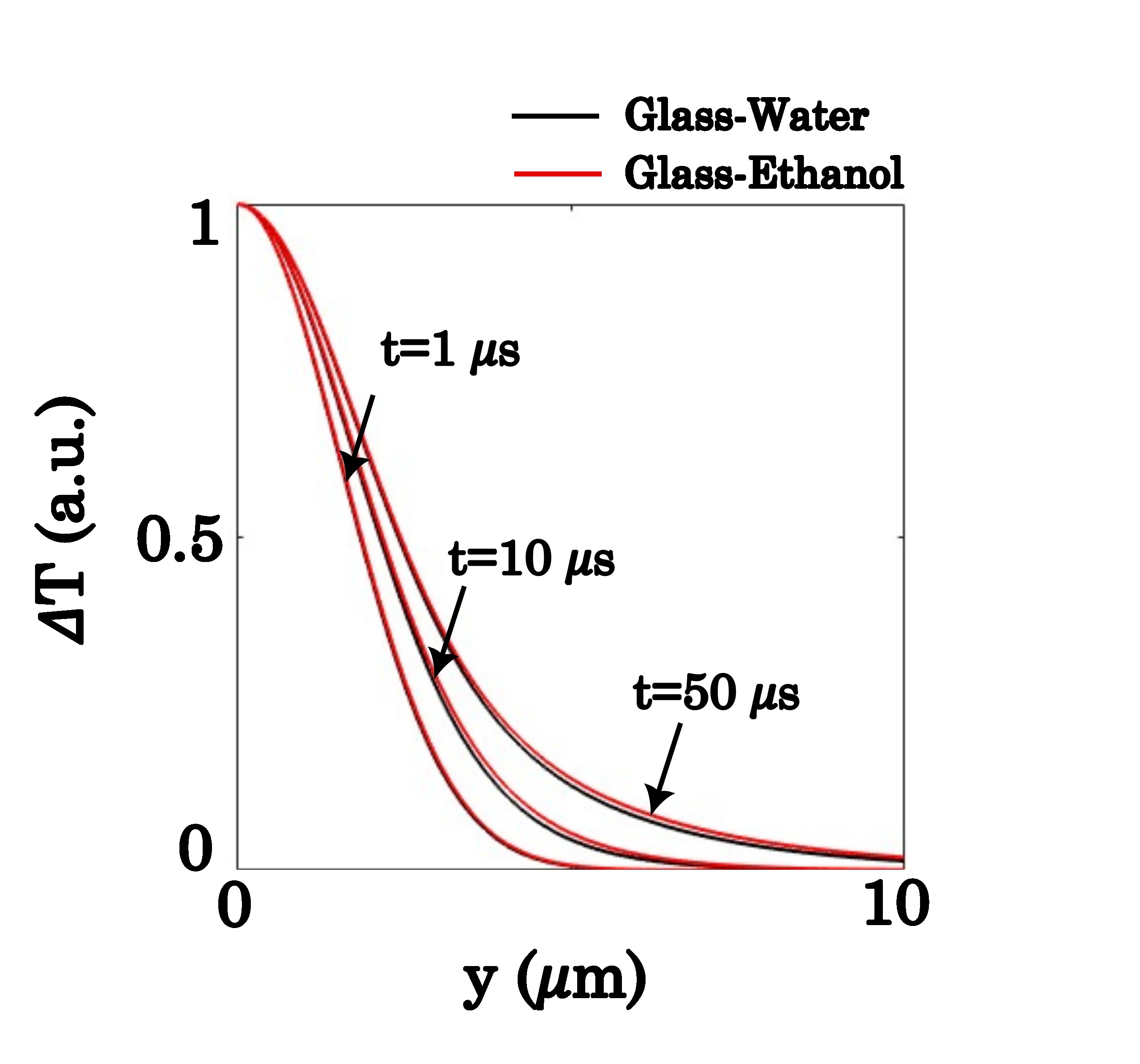}
  \caption{Comparison of normalized transverse temperature profiles Water--glass interface with Ethanol--glass interface obtained from COMSOL simulations. Results are shown for $t=1\,\mu\mathrm{s}$, $10\,\mathrm{ms}$, and $50\,\mathrm{ms}$. }
  \label{S4}
\end{figure}

Figure~\ref{S4} compares the transverse temperature profiles for the ethanol--glass and water--glass interfaces at early times ($t = 1~\mu\mathrm{s}$ to $50~\mu\mathrm{s}$). The thermal envelope expands more rapidly along the ethanol--glass interface than along the water--glass interface, reflecting the larger apparent interfacial diffusivity ($D_{\mathrm{app}}$) of the former. These results are consistent with the behavior described in the main manuscript and highlight the influence of interfacial thermal properties on transient heat spreading.

\end{document}